\documentstyle[psfig]{mn} 
\newif\ifAMStwofonts 
\def\kms{\thinspace\hbox{$\hbox{km}\thinspace\hbox{s}^{-1}$}}
\def\ergs{\thinspace\hbox{$\hbox{erg}\thinspace\hbox{s}^{-1}$}}
\def\ergscm{\thinspace\hbox{$\hbox{erg}\thinspace\hbox{s}^{-1}\thinspace\hbox{cm}^{-2}$}} 
\def\cc{\thinspace\hbox{$\hbox{cm}^{-3}$}}
\def\c2{\thinspace\hbox{$\hbox{cm}^{-2}$}}

\def\ha{\hbox{$\hbox{H}\alpha$}}  
\def\hb{\hbox{$\hbox{H}\beta$}}
\def\hg{\hbox{$\hbox{H}\gamma$}}
\def\hd{\hbox{$\hbox{H}\delta$}}
\def\msun{\thinspace\hbox{$\hbox{M}_{\odot}$}}
\def\msunyr{\thinspace\hbox{$\hbox{M}_{\odot}\thinspace\hbox{yr}^{-1}$}}
  
\def\kmsmpc{\thinspace\hbox{$\hbox{km}\thinspace\hbox{s}^{-1} \thinspace\hbox{Mpc}^{-1}$}}
\def\gapprox{$_>\atop{^\sim}$}     %Greater than over approximately%
\def\lapprox{$_<\atop{^\sim}$}        %Less than over approximately%
\def\si{$\sim$}

\def\arcdeg{\hbox{$^\circ$}}
\def\arcmin{\hbox{$^\prime$}}

\def\farcs{\hbox{$.\!\!^{\prime\prime}$}}

\def\hours{\hbox{$^{\rm h}$}}

\def\minutes{\hbox{$^{\rm m}$}}

\def\fseconds{\hbox{$.\!\!^{\rm s}$}}
\newcommand{\E}[1]{$\,10^{#1}$}                         %power of ten%
\newcommand{\ET}[1]{$\times 10^{#1}$}        %times power of ten%
\newcommand{\zp}[1]{\left( {#1} \right)}               %parentheses%
\newcommand{\smc}[1]{\,\small \sc {#1}}

\ifoldfss 
  \ifCUPmtlplainloaded \else 
    \NewTextAlphabet{textbfit} {cmbxti10} {} 
    \NewTextAlphabet{textbfss} {cmssbx10} {} 
    \NewMathAlphabet{mathbfit} {cmbxti10} {} % for math mode 
    \NewMathAlphabet{mathbfss} {cmssbx10} {} %  "   "    " 
  \fi 
  \ifAMStwofonts 
    \ifCUPmtlplainloaded \else 
      \NewSymbolFont{upmath} {eurm10} 
        \NewSymbolFont{AMSa} {msam10} 
      \NewMathSymbol{\upi}     {0}{upmath}{19} 
      \NewMathSymbol{\umu}     {0}{upmath}{16} 
      \NewMathSymbol{\upartial}{0}{upmath}{40} 
      \NewMathSymbol{\leqslant}{3}{AMSa}{36} 
      \NewMathSymbol{\geqslant}{3}{AMSa}{3E}

    \fi 
  \fi 
\fi % End of OFSS 
 
\ifnfssone 
  \newmathalphabet{\mathit} 
  \addtoversion{normal}{\mathit}{cmr}{m}{it} 
  \addtoversion{bold}{\mathit}{cmr}{bx}{it} 
  \newmathalphabet{\mathbfit} % math mode version of \textbfit{..} 
  \addtoversion{normal}{\mathbfit}{cmr}{bx}{it} 
  \addtoversion{bold}{\mathbfit}{cmr}{bx}{it} 
  \newmathalphabet{\mathbfss} % math mode version of \textbfss{..} 
  \addtoversion{normal}{\mathbfss}{cmss}{bx}{n} 
  \addtoversion{bold}{\mathbfss}{cmss}{bx}{n} 
  \ifAMStwofonts 
    \ifCUPmtlplainloaded \else 
      \UseAMStwoboldmath 
      \makeatletter 
      \new@mathgroup\upmath@group 
      \define@mathgroup\mv@normal\upmath@group{eur}{m}{n} 
      \define@mathgroup\mv@bold\upmath@group{eur}{b}{n} 
      \edef\UPM{\hexnumber\upmath@group} 
      \new@mathgroup\amsa@group 
      \define@mathgroup\mv@normal\amsa@group{msa}{m}{n} 
      \define@mathgroup\mv@bold\amsa@group{msa}{m}{n} 
      \edef\AMSa{\hexnumber\amsa@group} 
      \makeatother 
      \mathchardef\upi="0\UPM19 
      \mathchardef\umu="0\UPM16 
      \mathchardef\upartial="0\UPM40 
      \mathchardef\leqslant="3\AMSa36 
      \mathchardef\geqslant="3\AMSa3E 
    \fi 
  \fi 
\fi % End of NFSS release 1 
 
\ifnfsstwo 
  \DeclareMathAlphabet{\mathbfit}{OT1}{cmr}{bx}{it} 
  \SetMathAlphabet\mathbfit{bold}{OT1}{cmr}{bx}{it} 
  \DeclareMathAlphabet{\mathbfss}{OT1}{cmss}{bx}{n} 
  \SetMathAlphabet\mathbfss{bold}{OT1}{cmss}{bx}{n} 
  \ifAMStwofonts 
    \ifCUPmtlplainloaded \else 
      \DeclareSymbolFont{UPM}{U}{eur}{m}{n} 
      \SetSymbolFont{UPM}{bold}{U}{eur}{b}{n} 
      \DeclareSymbolFont{AMSa}{U}{msa}{m}{n} 
      \DeclareMathSymbol{\upi}{0}{UPM}{"19} 
      \DeclareMathSymbol{\umu}{0}{UPM}{"16} 
      \DeclareMathSymbol{\upartial}{0}{UPM}{"40} 
      \DeclareMathSymbol{\leqslant}{3}{AMSa}{"36} 
      \DeclareMathSymbol{\geqslant}{3}{AMSa}{"3E} 
    \fi 
  \fi 
\fi % End of NFSS release 2 
 
\ifCUPmtlplainloaded \else 
  \ifAMStwofonts \else % If no AMS fonts 
    \def\upi{\pi} 
    \def\umu{\mu} 
    \def\upartial{\partial} 
  \fi 
\fi 
 
\title[SN~1997eg]
{The circumstellar material around SN IIn 1997eg: \\
Another detection of Very Narrow
P~Cygni profile\thanks{Based on observations made with the 4.2-m
William Herschel Telescope, 
operated on the island of La Palma by the Isaac Newton
Group in the Spanish Observatorio del Roque de los Muchachos of the
Instituto de Astrof\'{\i}sica de Canarias. }}
\author[I. Salamanca, R.J. Terlevich \& G. Tenorio-Tagle]
       {Isabel Salamanca$^1$, Roberto J. Terlevich$^2$ and Guillermo
         Tenorio-Tagle$^3$ \\
$^1$Observatory of Leiden, Postbus 9513, NL-2300 RA Leiden, The Netherlands.\\
$^2$Institute of Astronomy, Madingley Road, Cambridge CB3 OHA, U.K. \\
$^3$Instituto Nacional de Astrof\'{\i}sica Optica y Electronica, AP 51, 72000 Puebla,
M\'exico. \\
}

\date{Accepted ... 
      Received ...; 
      in original form 2000 December 16} 
 
\pagerange{\pageref{firstpage}--\pageref{lastpage}} 
\pubyear{2000} 
 
\begin{document} 
 
\maketitle 
 
\label{firstpage} 
 
\begin{abstract} 
We report the detection of a very narrow P Cygni profile 
on top of the broad emission \ha\ and \hb\ lines of the Type IIn Supernova
1997eg. A similar feature
has been detected in SN~1997ab (Salamanca et al. 1998), SN~1998S
(Meikle \& Geballe 1998, Fassia et al. 2001) and SN~1995G (Filippenko \& Schlegel 1995). 
The detection of the narrow P Cygni profile indicates the
existence of a dense circumstellar material (CSM) into which the ejecta of
the supernova is expanding. From the analysis of the spectra of
SN~1997eg we deduce 
(a) that such CSM is very dense (n\gapprox 5\ET{7} \cc), (b) that has
a low expanding
velocity of about 160 \kms. The origin of such dense CSM can be either
a very dense progenitor wind (\.M \si \E{-2} \msunyr) or a circumstellar shell
product of the progenitor wind expanding into a high pressure
environment.
\end{abstract} 
 
 
\section{Introduction} 
 
SN~1997eg is a Type IIn supernova discovered on 1997 December
5 (Nakano \& Masakatsu 1997)
having an unfiltered CCD magnitude of 15.6. Its coordinates are R.A. $=$
13\hours 11\minutes 36\fseconds73 and DEC $=$ +22\arcdeg 55\arcmin 29\farcs4 (equinox 2000.0),
which is 4\farcs1 west and 33\farcs1 north of the center of the host galaxy
NGC~5012, a spiral galaxy with morphological type SAB(rs)c that host a
low luminosity AGN in its center, and is situated at 50 Mpc (Ho, Filippenko, Sargent 1997).
From the analysis of a spectrum taken 15 days later, Filippenko \& Barth (1997) report
that in the
optical range, SN~1997eg show the
typical features of Type IIn SN: absence of broad P Cygni profiles and,
instead, strong emission lines, notably \ha\ and \hb\
lines, on top of a
very blue continuum - like most Type IIn supernovae. The He{\smc I} 5876 \AA\ is very strong,
suggesting either a very high Helium abundance or a blend with the Na{\smc
  I} 5894 \AA\ blend. From the ratio of the [O{\smc III}]
4363/5007 lines they deduce the presence of a very dense circumstellar material (n
\gapprox \E{8} \cc). Besides, lines of very high excitation like
He{\smc II} 4886 \AA\ or [Fe{\smc X}] 6375 \AA\ are prominent (Filippenko \&
Barth 1997) indicating the presence of hard radiation.
Finally, SN~1997eg has been detected in radio 
with the Very Large Array (VLA). The flux measured was 0.52 $\pm$ 0.06
mJy at 3.6 cm on 1998 May 31 and 0.53 $\pm$ 0.12 on 1998 June 9 (Lacey
\& Weiler 1998).  
 
The exact date of explosion of SN~1997eg is not known, only that it was not seen on 1997
August 11 (\si 4 months before its discovery). We will adopt the
discovery date as the date of explosion. Therefore, our data were taken when the
supernova had an age of about 200 days. 

\section{The Observations}
On June 20 and July 16 1998, SN~1997eg was observed with the
Utrecht Echelle spectrograph (UES) and ISIS respectively at
the 4.2-m William Herschel Telescope (WHT) at ``El Roque de los
Muchachos'' Observatory (La Palma, Spain).
The echelle observations were done with a 2148$\times$2148 SITE CCD and the
79.0 lines/mm grating. The long slit ones were done with a 1024$\times$1024 Tek CCD
and the R158R grating for the red arm of ISIS and with a 2148$\times$4200
EEV CCD together with the grating R600B for the blue arm.
The observing log is given in Table~\ref{tab:log}. The data were
calibrated using the standard procedures within {\small \sc IRAF}: they were debiased,
trimmed, flat-fielded using a normalized flat field, and
calibrated in wavelength by using arc lamps of Th-Ar for the echelle
spectra and Ne-Ar for the long slit ones. Flux calibration was
performed by using the standard stars BD+33 2642 and BD+26 2606
respectively. The atmospheric extinction was applied using the
mean extinction curve for La Palma. Finally, the spectra were
corrected by redshift, thus all the wavelengths cited here are at rest frame.

\begin{figure*}
\vspace{0cm}  % amount of vertical space needed
\centering
\psfig{figure=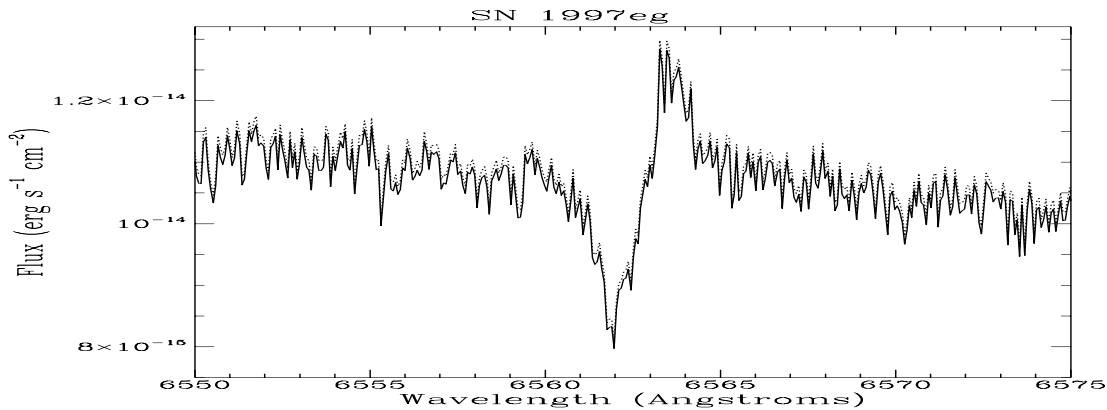,height=6cm,width=15cm}
    \caption{The narrow P~Cygni is not affected by the galaxy
      subtraction, due to the extended \ha\ and \hb\ emission. In this
      figure we show two spectra, one extracted with background subtraction
      (solid line) and the other with no background subtraction (dotted line). As we can see
      there is practically no difference among them.}
    \label{fig:checkPC}
\end{figure*}

\begin{table*}
\begin{minipage}[ctb]{120mm}
\caption{Log of spectroscopic observations of
SN~1997eg. The ``+day'' 
is the number of days elapsed since explosion (1997 December 5).}
\label{tab:log}
\begin{tabular}{|lccccc|}
\hline
Date   & age     & Instrument  & Slit  & Spect. range  & Spect. resol.$^a$ \\
       & (days)  &           & (arcsec) & (\AA)    & (\AA\ / \kms)  \\
\hline
\hline
20 June & +198   & UES  & 1.2  & 4400 - 8977 &  0.10 - 0.19 / 6.5 \kms \\
16 July & +224   & ISIS Blue arm& 2.0  & 3700 - 5480 & 3 / 185 \kms  \\
16 July & +224   & ISIS Red arm & 2.0  & 5550 - 8500 &  11.8 / 540 \kms \\
\hline 
\end{tabular}
$^a$ As measured from the FWHM of the arc lines\\
\end{minipage}
\end{table*}
\begin{figure*}
\vspace{0cm}  % amount of vertical space needed
\hbox{\hspace{-0.2cm}\psfig{figure=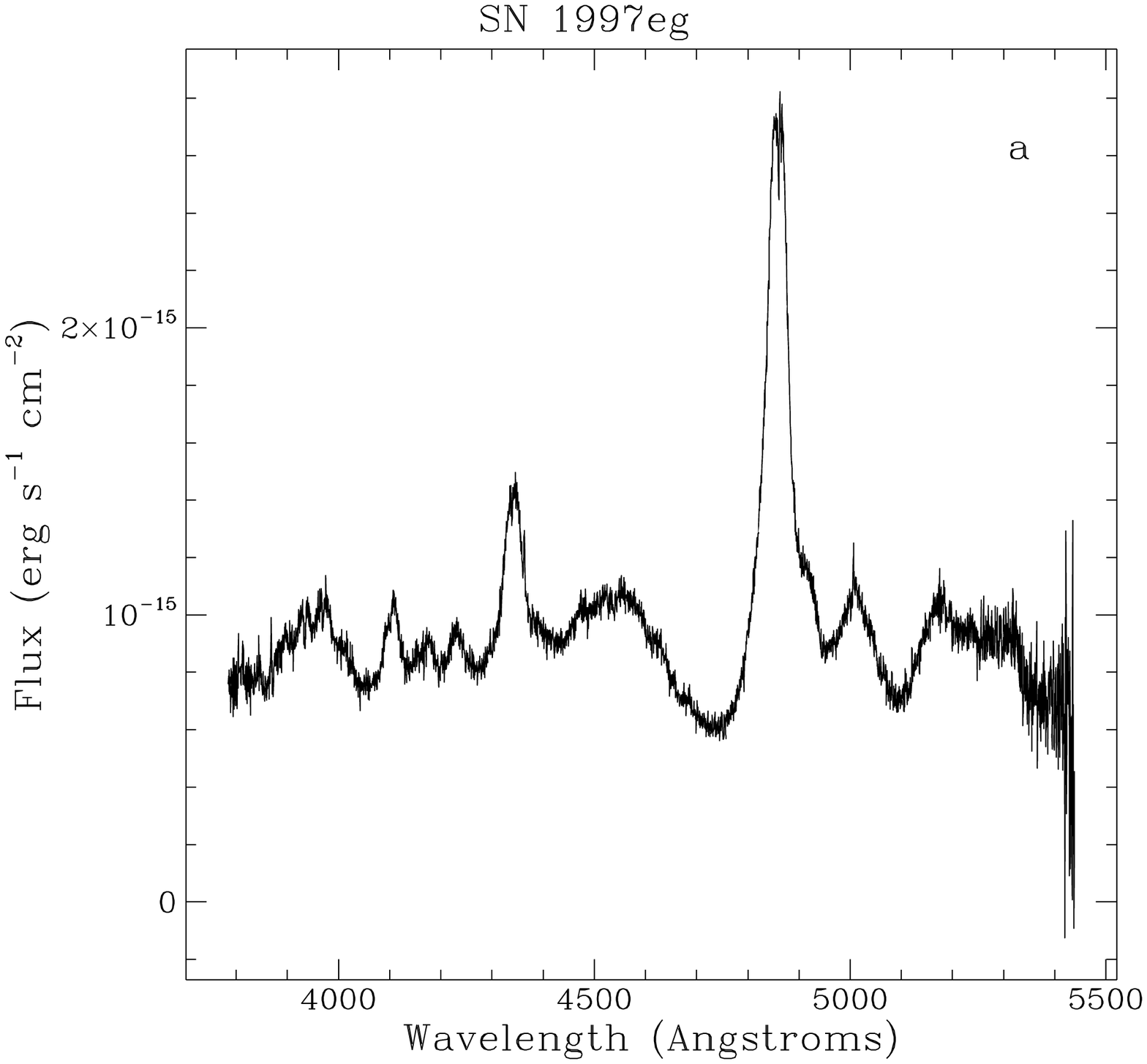,height=8cm,width=8cm}\hspace{0.8cm}
\psfig{figure=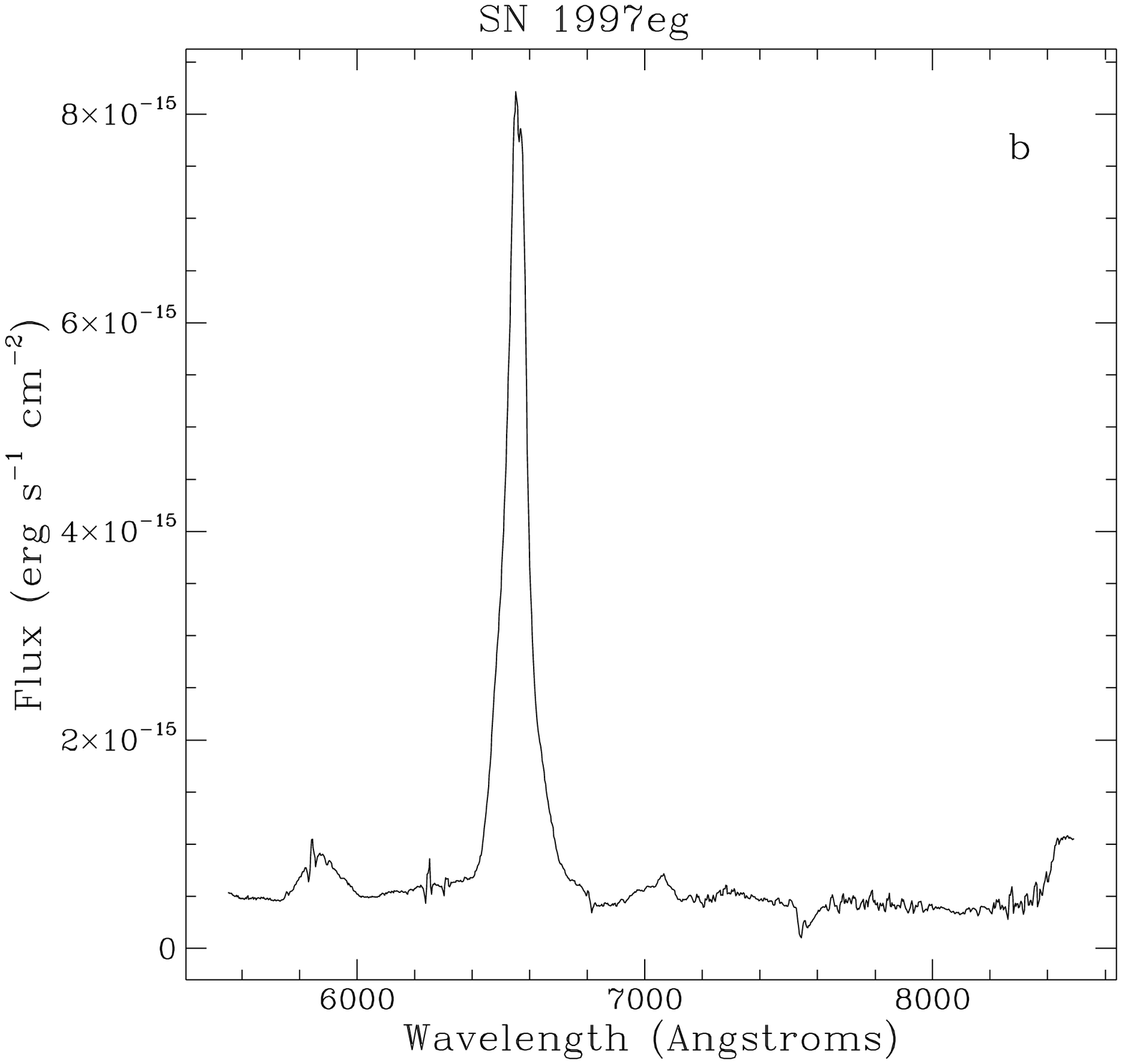,height=8cm,width=8cm}}
\vspace{0cm}
\hbox{\hspace{-0.2cm}\psfig{figure=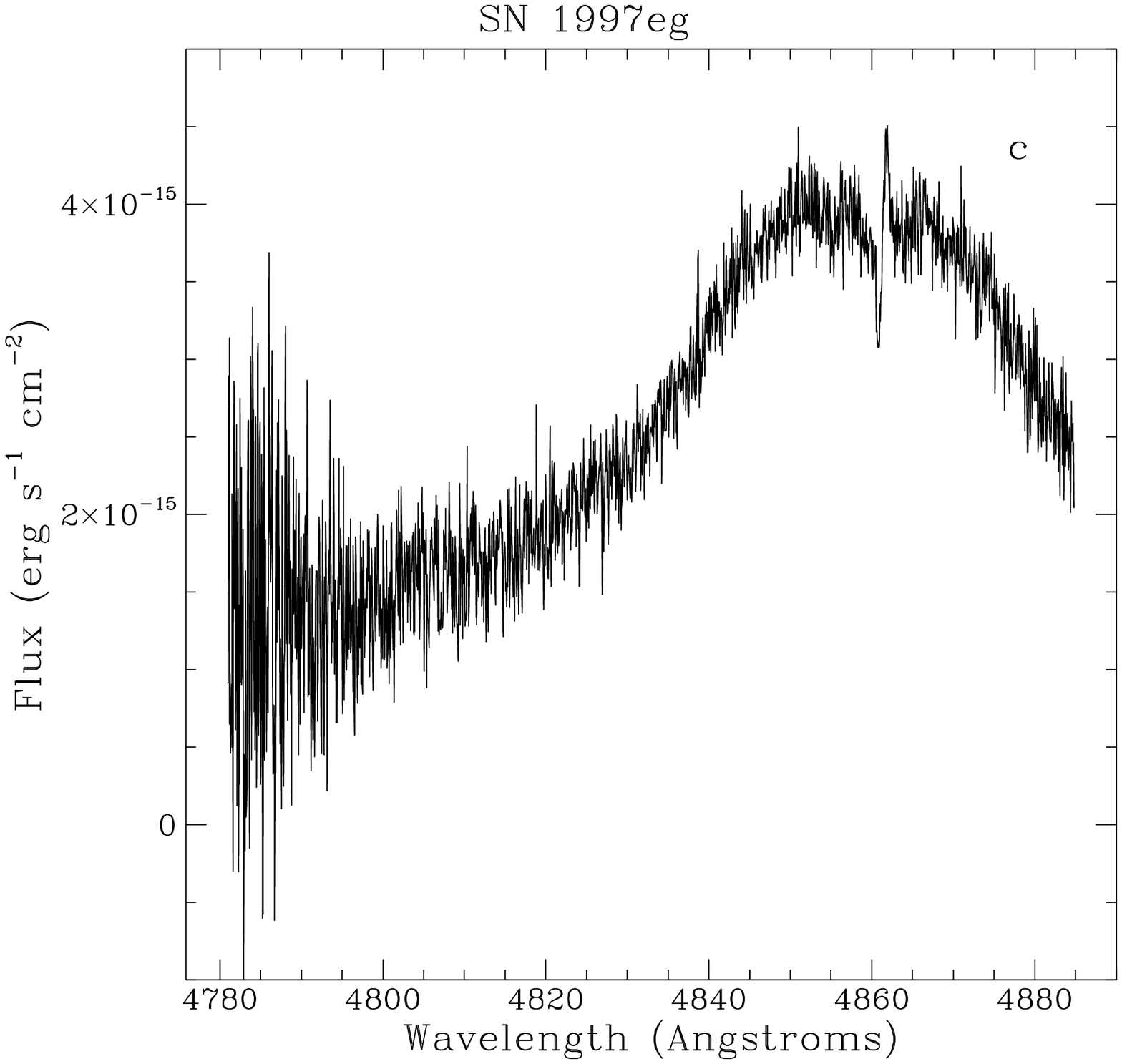,height=8cm,width=8cm}\hspace{0.8cm}
\psfig{figure=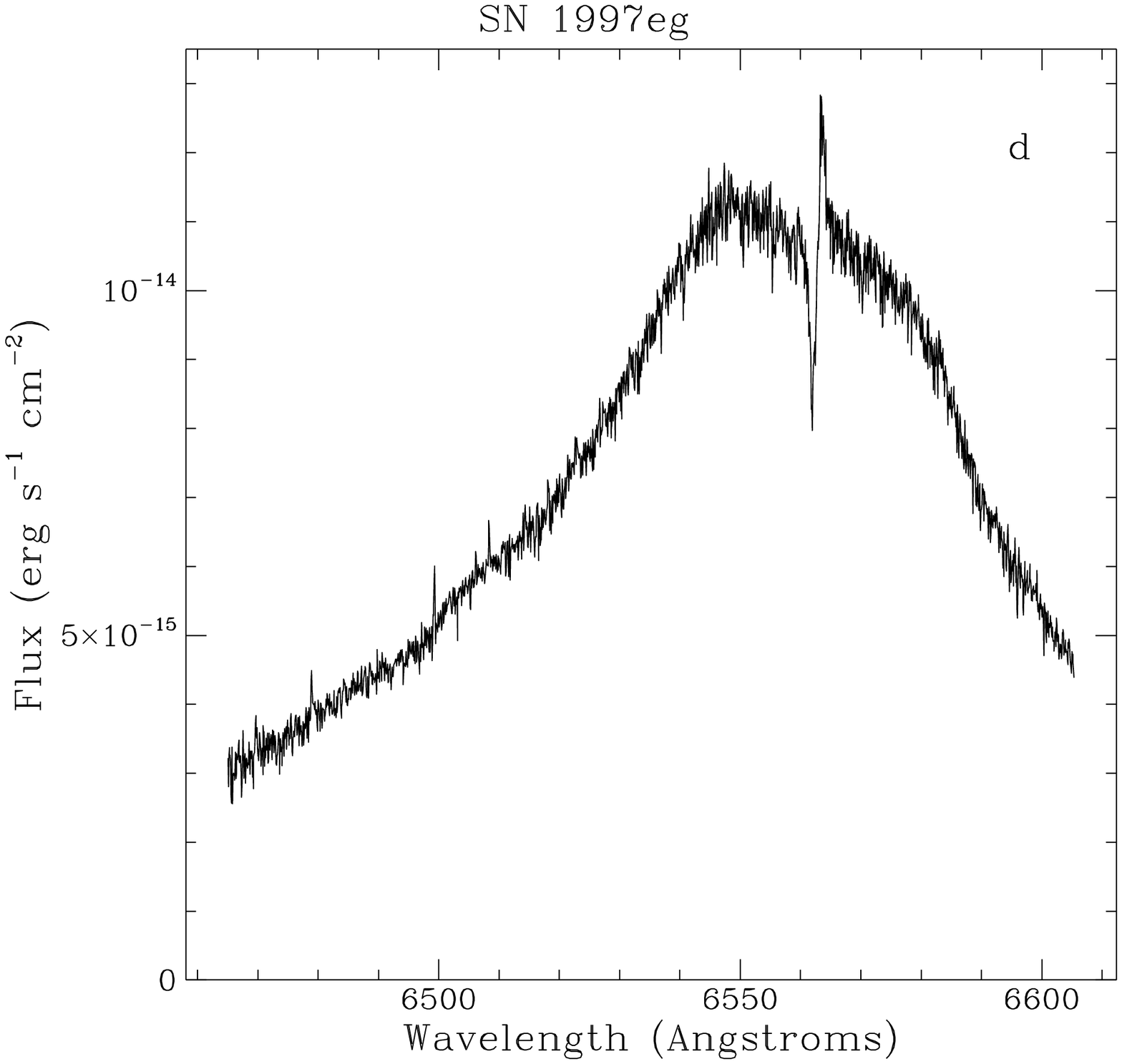,height=8cm,width=8cm}}
\caption{Spectra of SN1997eg. On the top panels, the
long slit spectra taken with ISIS, (a) \hb\ and (b) \ha, and on the bottom the echelle
spectra (only the
order with \hb\ (c) and \ha\ (d) are shown). The flux is in units of
\ergscm\ and the wavelength is in Angstroms.}
\label{fig:longslit}
\end{figure*}

\begin{figure*}
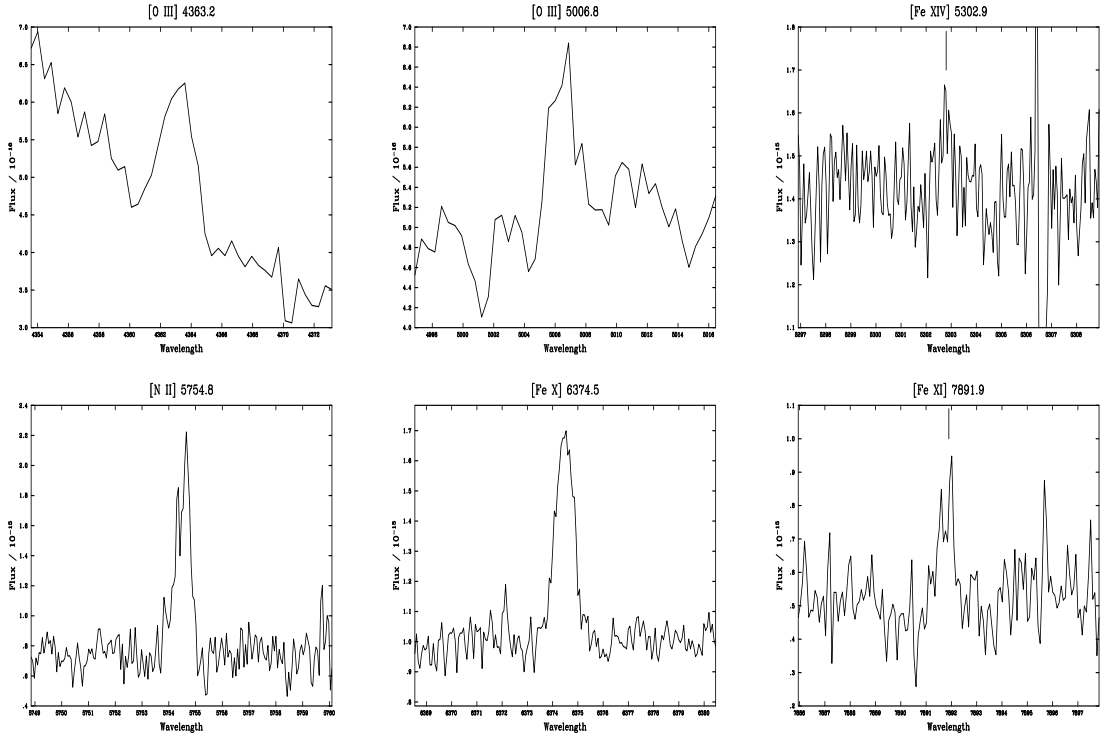

\vspace{0cm}  % amount of vertical space needed
\hbox{\hspace{0.2cm}\psfig{figure=OIII_4363_new.eps,height=5cm,width=5cm,angle=-90}\hspace{0cm}
\psfig{figure=OIII_5007_new.eps,height=5cm,width=5cm,angle=-90}\hspace{0cm}
\psfig{figure=FeXIV_5303_new.eps,height=5cm,width=5cm,angle=-90}}
\vspace{0cm}
\hbox{\hspace{0.2cm}\psfig{figure=NII_5755_new.eps,height=5cm,width=5cm,angle=-90}\hspace{0cm}
\psfig{figure=FeX_6375_new.eps,height=5cm,width=5cm,angle=-90}\hspace{0cm}
\psfig{figure=FeXI_7892_new.eps,height=5cm,width=5cm,angle=-90}}
\caption{The most prominent narrow lines that are seen in the echelle
  spectra, ordered by increasing wavelength. The wavelength range is
 12 \AA\ except for [O{\smc III}] 4363 \AA\
 and [O{\smc III}] 5007 \AA\ which is 20 \AA. }
\label{fig:narrow_prominent}
\end{figure*}

\section{Analysis}

\subsection{Redshift and distance of SN~1997eg}

The redshift, and therefore the heliocentric radial velocity and
distance to SN~1997eg can be determined from the narrow lines. We take
the echelle spectra because of its higher spectral resolution. We
choose the narrow lines [Fe{\smc XIV}] 5303 \AA, [N{\smc II}] 5755 \AA\ and [Fe{\smc X}] 6375
\AA. These are among the lines with better signal-to-noise ratio in the echelle spectra and that are 
clearly associated with the circumstellar material of SN~1997eg (see section 3.5 and Fig.~7). The line 
[Fe{\smc XI}] 7892 \AA\ was discarded because of its double peak (see Fig.~3).
The average redshift is
0.00837 $\pm$ 0.00004, equivalent to 2511 $\pm$ 12 \kms. The heliocentric velocity is
therefore 2485 $\pm$ 12 \kms.  
If we adopt a Hubble constant of H$_o$ = 50 \kmsmpc\ then the distance to
SN~1997eg is 50 Mpc. 
Since the emission line in a P~Cygni profile is
partially absorbed it is difficult to position exactly the center of
the profile. However, after having corrected the echelle spectra by
redshift, the center of the \ha\ P~Cygni is at \si 6562.9 \AA. 

\subsection{Contamination by the host galaxy}
\label{sec:contamination}
The extended \ha\ and \hb\ emission of the host galaxy, NGC~5012, is clearly visible on the
2D images of SN~1997eg. We have checked how much of the narrow P~Cygni
profile can be due to an effect of background subtraction.
For that purpose we have extracted the echelle spectrum
with and without background subtraction, and we have compared both
results. In Fig.~\ref{fig:checkPC} we can see that the \ha\ P~Cygni in both spectra
- with and without background subtraction - is almost equal. Therefore the extended \ha\
emission does not play a significant role in the P~Cygni profile.

\subsection{Line identification}

\begin{table*}
  \begin{minipage}[ctb]{155mm}
\caption{List of emission lines seen in the  spectra of SN~1997eg. The first
column is the observed (redshifted) wavelength in Angstroms. The second column is 
the line identification, the third is the full width at half maximum
(corrected by the instrumental response) the fourth is the flux in
units of \E{-16} \ergscm, the fifth indicates if the line is seen in
the echelle spectra (UES) or the long-lsit one (ISIS), and the last
column is the Ionisation Potential of the line.}
\label{tab:lineas}
\begin{tabular}{llllll}
\hline
Wavelength & Identification & FWHM & Flux     & Spectrum & Ion. Potential \\
 \AA       &                & \kms  & \ergscm &          &  eV \\
\hline
\hline
3868.3      & [Ne{\smc III}] 3868.7 & unresolved   & 5.0      & ISIS  & 40\\
\si 4107    & \hd           & \si 2390     & 76       & ISIS  & 13.6 \\
\si 4343    & \hg           & \si 3510     & \si 200  & ISIS$^a$ & 13.6  \\
4339.2      & \hg           & unresolved   & \si -2.2 & ISIS$^b$  & 13.6 \\
4363.4      & [O{\smc III}] 4363.2  & unresolved   & 5.3      & ISIS  & 35.1 \\
4486.43     & [Fe{\smc II}] 4486.23  & \si 35    & 7        & UES   & 7.9 \\
4609.04     &               & 26           & 7.8      & UES   & - \\
4658.26     & Fe{\smc II} 4658.0 or [Fe{\smc III}] 4658.1 & 8 & 1.5 & UES & 7.9 or 16.18 \\
4739.50     & Mg{\smc II} 4739.6  & unresolved   & 1.6      & UES    & 7.6 \\ 
4838.69     & 4838.7        & unresolved    & 0.9     & UES$^c$ & - \\
4861.6     & \hb           & 3200 (core)   & 880     & ISIS \& UES$^d$  & 13.6 \\
4860.81    & H$\beta$      & 40            & -5.9    & UES \& ISIS &13.6 \\
4861.86    & H$\beta$      & 36            & 3.2     & UES \& ISIS &13.6 \\
4899.72    & 4899.6        & unresolved    & 2.8     & UES$^c$ & - \\
4906.29    & Fe{\smc IV} 4906.2  & unresolved    & 2.2     & UES     & 30.65  \\
4958.80    & [O{\smc III}] 4958.9 &\si 30        & 2.0     & UES$^e$ & 35.1   \\
4977.84    & 4977.6        & \si 23       & 1.5     & UES$^c$ & - \\
4988.36    &               & unresolved  & 0.5      & UES     & -  \\
5006.81    & [O{\smc III}] 5006.8 & \si 40        & 3.8    & ISIS \& UES$^f$ & 35.1 \\
5040.99    & Si{\smc II} 5041.1 & unresolved & 0.7 + 0.9   & UES    & 8.15 \\
5077.16    & [Fe{\smc II}] 5076.6  & unresolved  & 0.6     & UES    & 7.9  \\
5159.60    & [Fe{\smc II}] 5158.8  & \si 6.5$^g$      & 0.8  & UES    & 7.9  \\
5176.03    & N{\smc II} 5175.9     & \si 6.5$^g$  & 0.9      & UES    & 14.5 \\
5270.59    & [Fe{\smc II}I] 5270.4 & \si 25      & 1.3     & UES    & 16.18 \\
5302.78    & [Fe{\smc XIV}] 5302.9 & \si 38      & 2.0     & UES$^h$ & 361.0 \\
5429.18    & [Fe{\smc VI}] 5428.6 or Fe I 5429.7& 14 & 0.8 & UES   & 75.5 or 0\\
5438.97    &                  & 30         & 0.8    & UES   & -\\
5440.18    & [Fe {\smc II}] 5440.5 & unresolved    & 0.3   & UES   & 7.9 \\
5525.72    & Fe {\smc II} 5525.1    & 36+unresolv.  & 2.0  & UES   & 7.9 \\
5580.59    & [Fe {\smc II}] 5580.8  & 8           & 0.9     & UES  & 7.9 \\
5700.96    &                 & 14          & 1.0     & UES & - \\
5754.65    & [N {\smc II}] 5754.8  & \si 34      & 9.4     & UES \& ISIS & 14.5\\ 
5843.5     &                & 400         & 18      & ISIS & - \\
\si 5875   & He{\smc I} 5876         & 7500        & 670   & ISIS & 24.6 \\
5875.62    & He{\smc I} 5875.6 & \si 10 + 30-40   & -      & UES &  24.6 \\
5875.98    & He{\smc I} 5876.0 &   -               & 2.3 (+)   & UES$^i$ & 24.6 \\
5949.15    &               & 11           & 1.1     & UES &  - \\
6374.48    & [Fe{\smc X}] 6374.5  & 42           & 7.6     & UES$^h$ & 235.0 \\
6506.15    &               & unresolved   & 0.8     & UES &  -\\
\si 6551.2  & \ha          & 3800         & 9077  & ISIS \& UES$^j$ & 13.6 \\  
6561.96     & \ha          & 53           & -36.4  & UES \& ISIS$^b$ &13.6  \\
\si 6563.5  & \ha          & 40           & 14.0   & UES$^k$  & 13.6 \\
7065.5     & He{\smc I} 7065   & \si 2750  + \si 6000 & 24.6 & ISIS$^l$ &24.6\\
7468.33    & N{\smc I} 7468.3   & 12            & 2.5    & UES   & 0 \\
7891.98    & [Fe{\smc XI}] 7891.9 & 31          & 3.3     & UES$^h$   & 262.1 \\
\hline
\hline
\end{tabular}
$^a$Flat topped\\
$^b$The narrow absorption of the P Cygni profile\\
$^c$Seen on symbiotic stars \\
$^d$Flat toped \\
$^e$Very noisy \\
$^f$UES marginal \\
$^g$Just resolved \\
$^h$Coronal line \\
$^i$This line and the previous are not totally separated \\
$^j$The strongest line \\
$^k$Narrow emission peak not well defined \\
$^l$Broad component blue-shifted \\
\end{minipage}
\end{table*}

The most prominent lines are the Balmer series, \hd, \hg, \hb\ and
\ha\ (see Fig.~\ref{fig:longslit})
together with Fe{\smc II} emission.
Other lines detected in the long slit spectra are the Oxygen lines, 
[O{\smc III}] 4363 \AA\ and [O{\smc III}] 5007 \AA, the lines of Helium 
He{\smc I} 5876 \AA\ and He{\smc I} 7065 \AA, and 
the blue side of the infrared Ca{\smc II} triplet at \si 8500 \AA.
In addition,  we identify in the echelle spectra several high excitation iron lines: [Fe{\smc
XIV}] 5303 \AA, [Fe{\smc X}] 6375 \AA\ and [Fe{\smc XI}] 7892 \AA\ (see
Fig.~\ref{fig:narrow_prominent}) as well as several low excitation lines of Fe{\smc II},
[Fe{\smc II}] and [Fe{\smc III}]. In
Table~\ref{tab:lineas} we give a list of the emission lines identified
in both, the long slit and echelle spectra. The Full Width at Half
Maximum (FWHM) of the narrow lines is the one measured on the echelle
spectra whenever possible, and it has been corrected by the
instrumental resolution.

Finally, 
we must as well draw the attention to those lines that are {\it not}
detected, like [N{\smc II}] 6583 \AA, [O{\smc I}] 6300,6364 \AA\
or the coronal lines [Ar{\smc X}] 5533 \AA,
[Ca{\smc V}] 5309 \AA\ and [Fe{\smc XI}] 7889 \AA \footnote{The coronal
line [Fe{\smc VII}] 6087 \AA\ falls in between two orders of the
echelle spectra and therefore nothing can be said about its detection.}. 

\subsection{Balmer lines profiles}
\label{sec:profiles}

\begin{figure*}
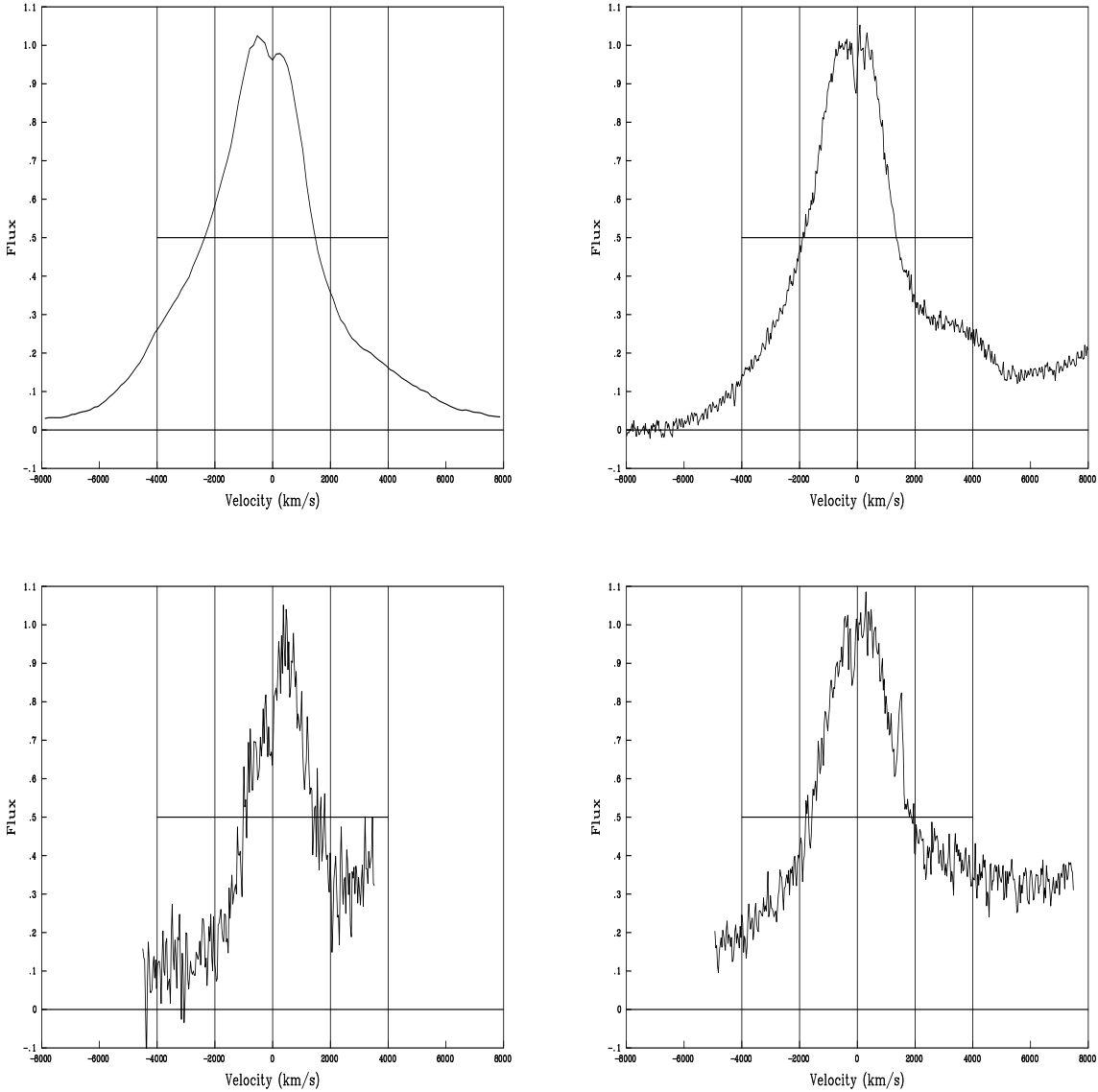

  \centering
%  \begin{minipage}{160mm}
\vspace{0cm}  % amount of vertical space needed
\hbox{\hspace{-0.5cm}\psfig{figure=profileHa_Core.eps,height=8cm,width=8cm,angle=-90}\hspace{0cm}  
\psfig{figure=profileHb_Core.eps,height=8cm,width=8cm,angle=-90}}
\vspace{0cm}
\hbox{\hspace{-0.5cm}\psfig{figure=profileHd_Core.eps,height=8cm,width=8cm,angle=-90}\hspace{0cm}
\psfig{figure=profileHg_Core.eps,height=8cm,width=8cm,angle=-90}}
\caption{Velocity profile of the broad Balmer lines. The
    continuum has been subtracted and the top of the lines has been
    normalized to one. The vertical lines correspond to velocities
    equal to 0, $\pm$ 2000 and $\pm$ 4000 \kms$\!$. The horizontal line
    marks the point of half maximum. The lines are (from left to right
    and from top to bottom) \ha, \hb, \hg\ and \hd.}
\label{fig:Bvel}
%\end{minipage}
\end{figure*}

We show in Fig.~\ref{fig:Bvel} the profile of the four strongest
Balmer lines, \ha, \hb, \hd\ and \hg.
The continuum has been  
fitted and subtracted using a straight line and the maximum  
amplitude of the line has been set to one. For each line, 
the zero velocity corresponds to its rest-frame wavelength. 
Note that this corresponds to the central part of the narrow P Cygni profile.
All the Full Width at Half Maximum (FWHM) given in this section have
been corrected by instrumental response. 

The most evident characteristic of the  broad Balmer lines is that their profiles are asymmetric and
flat topped. 

\begin{figure*}
\vspace{0cm}  % amount of vertical space needed
\hbox{\hspace{-0.5cm}\psfig{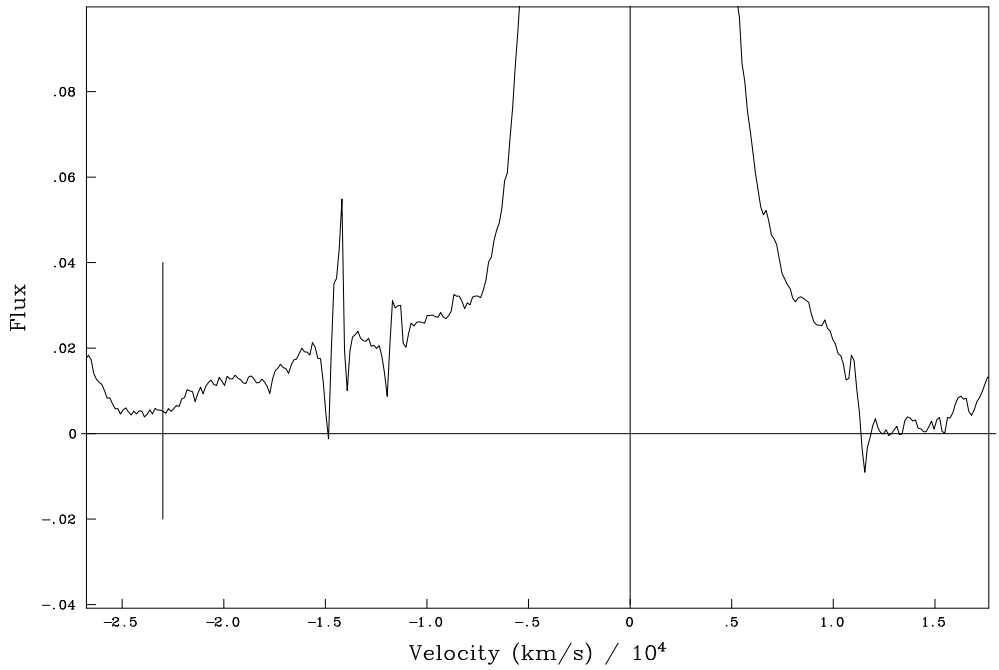}}
\caption{Velocity profile of the wings of the \ha\ line. The
    continuum has been subtracted and the top of the line has been
    normalized to one. The vertical line marks the approximated end of
    the very extended wing, \si 23000 \kms$\!$.}
\label{fig:Havel}
\end{figure*}

The \ha\ line seems composed of a central core and broad wings and has
the red wing less extended that the blue wing. This
is probably due to self-absorption. The
FWHM is \si 3800 \kms\ and
The Full Width at 10\% intensity is \si 11000 \kms$\!$. \ha\ has also a very
extended and weak blue wing
extending up to \si 23000 \kms\ (see Fig.~\ref{fig:Havel}). A similar
extended blue wing was
detected in SN~1997ab in the early spectra. In fact, overall, the \ha\
profile in both supernovas, SN~1997ab and SN~1997eg,  is remarkably similar, despite the probable
difference of about 200 days in their age. The narrow P Cygni
absorption is also detected in the long slit spectrum  
but not resolved. 

The \hb\ line is also asymmetric, and the red wing less
extended than the blue one. However this effect  
is less pronounced than in \ha. The FWHM of \hb\ is \si
3200 \kms. Unfortunately, the red wing of \hb\ is blended with Fe{\smc II}
emission, and nothing 
can be said about the extension of the wings.
The core appears as flat-topped, and the narrow P Cygni line is
detected, although it is not resolved in the
low dispersion data.

The next Balmer line, \hg, has a core of about 3500 \kms\ FWHM, and is 
clearly flat-topped. The absorption component of the narrow P Cygni is
detected, but the emission (if any) is lost in the noise. 
The \hd\ line is also present, although weaker and consequently with
lower signal-to-noise ratio. Nevertheless, the absorption component of
the narrow P Cygni is detected. The core of the broad line has a FWHM
\si 2400 \kms$\!$. Contrary to \ha\ and \hb, the red side is more extended.
The extension of the wings in these two lines is difficult to determine,
because they are blended with Fe{\smc II} emission.

\begin{figure*}
\vspace{0cm}  % amount of vertical space needed
\hbox{\hspace{-0.5cm}\psfig{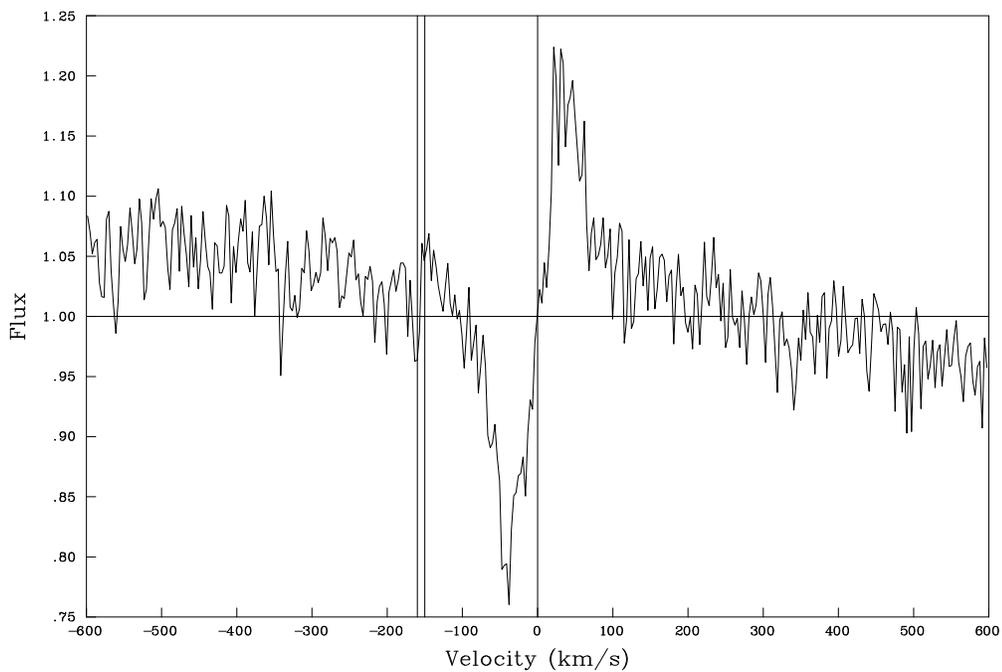}}
\caption{Velocity profile of the narrow P Cygni \ha\ line. The top of
  the broad line has been
    normalized to one. The vertical lines marks the center of the P
    Cygny and the approximated blue end of
    the absorption \si 150 - 160 \kms$\!$.}
\label{fig:PCvel}
\end{figure*}

In the high dispersion echelle spectra, we see
that effectively both \ha\ and \hb\ lines are flat-topped (the rest of
the Balmer series are outside the wavelength coverage of the echelle
data). We can also see that the 
narrow, unresolved component becomes a clear P~Cygni profile. The
absorption component seems to be somewhat broader than the emission
part. From the blue wing of the absorption we deduce that the material
emitting such a profile is expanding at 160 \kms\
(see Fig.~\ref{fig:PCvel}).

\subsection{The narrow lines}

Another important observational characteristic on SN~1997eg is the
presence of very narrow lines (FWHM between \si 7 and 40 \kms) with a
wide range of ionization potentials up to 360 eV.
The most prominent ones are shown in
Fig.~\ref{fig:narrow_prominent}.
This wide range of ionization energies 
implies the existence of an ionizing continuum that extends at
least between this values. In other words, SN~1997eg must be producing
soft X rays. 

\begin{figure*}
\vspace{0cm}  % amount of vertical space needed
\hbox{\hspace{-0.5cm}\psfig{figure=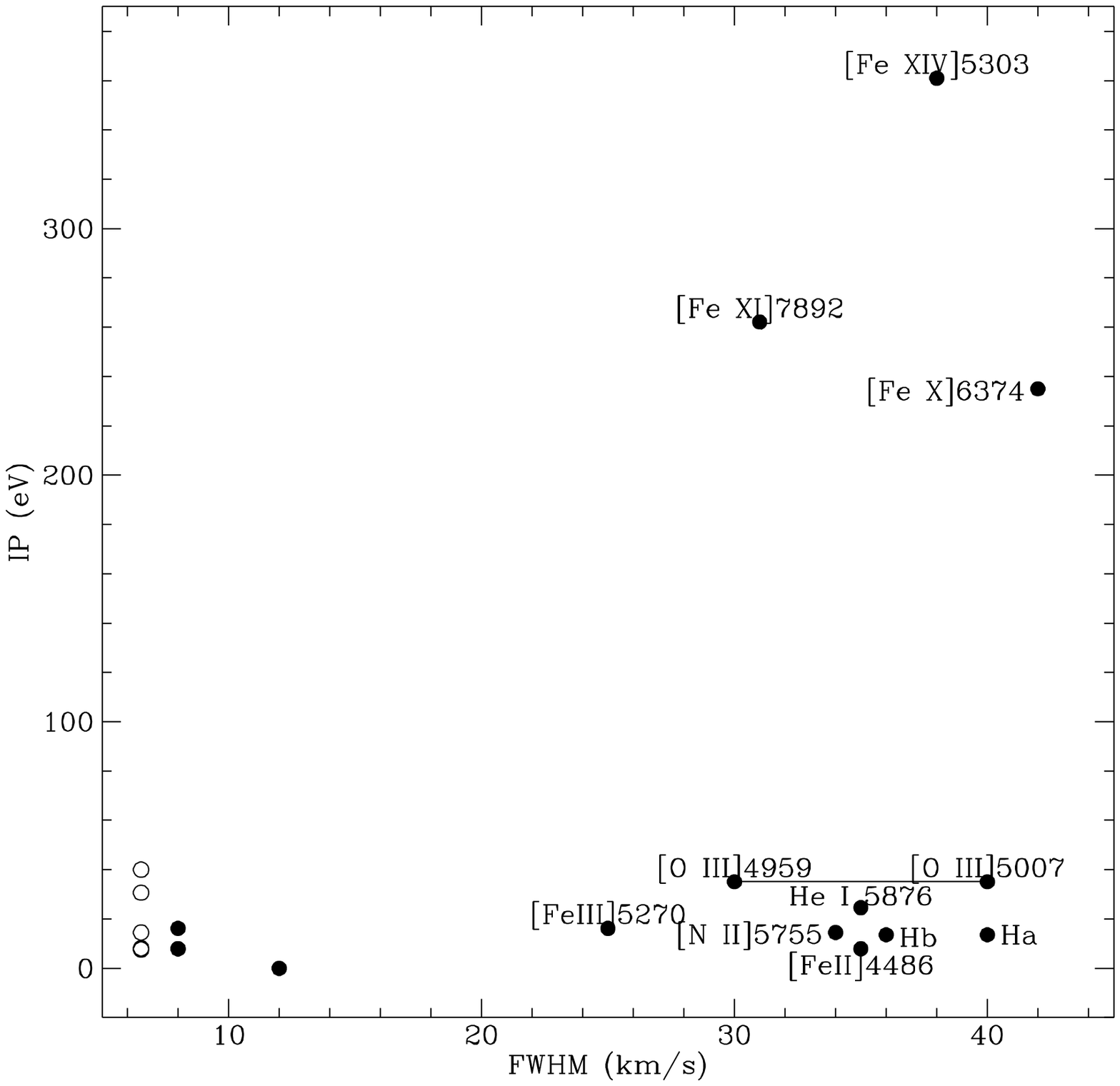,height=12cm,width=12cm}}
\caption{Ionization potential versus the FWHM of the narrow
  lines. Only those lines in the echelle data have been used. As well,
  those lines whose identification was not clear have been removed
  from the plot. The open circles at the bottom left correspond to
  those lines which were unresolved.}
\label{fig:ipvsfw}
\end{figure*}

If we plot the ionization potential (IP) of each narrow line versus its
FWHM (Fig.~\ref{fig:ipvsfw}) then it becomes clear that there are two
different groups of lines. One set, with low IP (between 0 and 40 eV)
and small FWHM (between 6 and 14 \kms$\!$). The other group has a much
broader range of IP (between 8 and 361 eV), and a larger FWHM (between
25 and 40 \kms$\!$).
This second group is probably related to the dense CSM that is surrounding the
supernova, while the first may be related to the H II region where
SN~1997eg exploded.

Finally, we must draw the attention to the fact that there are no broad
forbidden lines.

\section{Interpretation: A dense circumstellar material}
\label{sec:theory}

Type IIn Supernovae are associated with strong 
radiative shocks produced by the interaction of the fast
SN ejecta with a dense circumstellar material (Shull 1980, Wheeler, Mazurek, Sivaramakrishnan
1980, Chugai 1991, Chevalier \& Fransson 1994, Terlevich et al. 1992, 1995). 
When a supernova explosion occurs, two shocks are produced: the leading
shock ($v_{shock} \sim 10^4$ \kms), that encounters and heats the CSM up to a temperature 
of \si \E9 K, and the reverse shock ($v_{shock} \sim 10^3$ \kms), that decelerates and thermalizes the
ejecta to temperatures of about \E7 K. In the standard case, when the SN explosion occurs in a medium of 
density \si 1 \cc, both shocks are adiabatic, and remain so for thousands
of years. However, for very high densities (n \gapprox \E{7} \cc) the
adiabatic phase 
is almost nonexistent, and the shocks become radiative a few months after
the explosion, when the ejecta is still moving at high speeds (several times 1000 \kms), and when the
size of the remnant is less than 0.01 pc (hence the name of ``compact SN remnants'' given
by Terlevich et al. 1992, 1995).
When this happens, we have the phenomena called ``catastrophic
cooling'', which means that the time scale for cooling is much lower
than the time scale for pressure adjustment. In other words, there is a fraction of
the shocked gas that cools and loses pressure too rapidly for the gas to
adjust, and this generates secondary shocks that compresses the gas
into two thin shells (one for each shock front, leading and forward). 
These thin layers have a density of \si \E{9-13} \cc, a temperature of \si \E{4} K
and a velocity of few $\times$ 1000 \kms\ (Terlevich et al. 1992).

Theoretical calculations show that if the CSM has a constant density,
both the leading and reverse shocks become radiative 
(see Terlevich et al. 1992, 
1995). However, if the density of the surrounding media decreases as $r^{-2}$
as is expected from a wind, the leading shock may stay adiabatic,
and only the less luminous reverse shock becomes radiative (Chevalier \& Fransson
1994).
In any case, when either both or one shock becomes radiative, they will
emit copious amounts of UV and X-ray photons which will ionize 
the freely expanding ejecta, the dense and fast thin shell and the 
surrounding CSM. The CSM is heated to a temperature of \si \E{6-7} K
to decay in a few months to \si \E{5-6} K. 

The thin shell and the free expanding ejecta are 
responsible for the broad lines, such as \ha\ and \hb, as well as for the blue
continuum. On the other hand, the narrow emission lines 
(including the P Cygni) are emitted in the dense circumstellar
material. 
Therefore, we can distinguish two main emitting regions: the
circumstellar medium around SN~1997eg and the shocked material. The
evolution of the latter is dependent on the physical
conditions of the first, mainly its density. 

The fact that the broad \ha\ and \hb\ lines are flat-topped confirms that they are
emitted in an expanding thin shell. On the other hand,
the presence of the narrow P~Cygni profile proves the existence of
a slowly expanding, dense material into which the ejecta of the
supernova is expanding. 
Further observational evidence for the existence of such dense
material comes from Fig.~\ref{fig:ipvsfw}. There is a group of lines
which have similar FWHM (between 30 and 40 \kms) but quite large range
of IP. All the forbidden lines in this group have large critical
densities (\gapprox \E{5} \cc).

\begin{figure*}
\vspace{0cm}  % amount of vertical space needed
\hbox{\hspace{-0.2cm}\psfig{figure=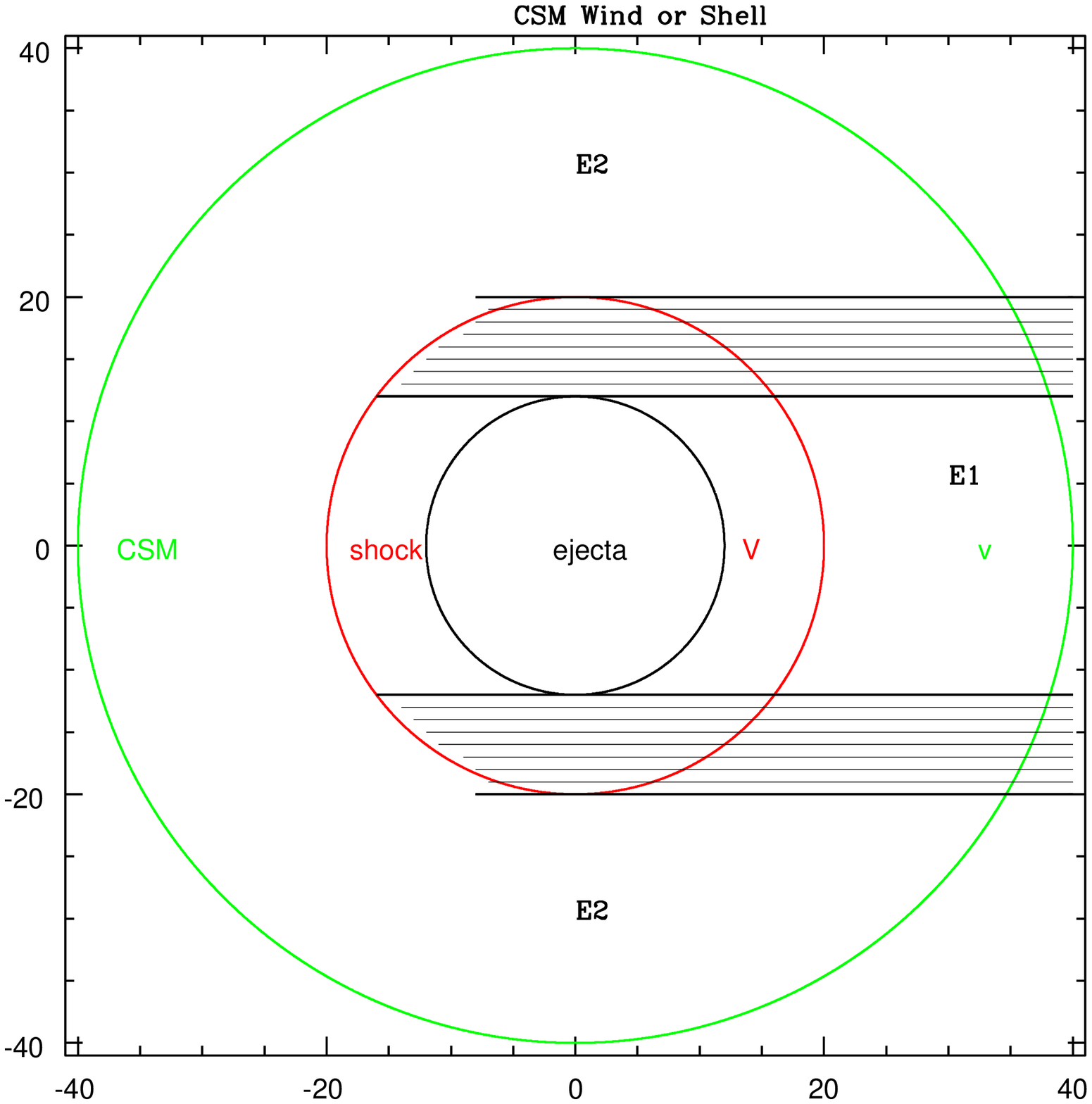,height=8cm,width=8cm}}
\caption{The different layers in both cases, CSW and CSSh. The
  observer is situated on the right side. The absorption would come
  from the shadowed area, where the projected velocities of the ejecta and the CSM
are equal. This,  seen from the front would form a
  narrow ring. The emission would come from the areas labeled E1 and
  E2.}
\label{fig:circulos}
\end{figure*}

\subsection{Observed parameters}
\label{sec:input}

The following physical parameters can be deduced from the observations:

\begin{description}
\item The luminosity of the broad Balmer lines: \\
$L_{H\alpha}^{B}=2.8$\ET{41} erg s$^{-1}$ \\
$L_{H\beta}^{B}=4.5$\ET{40} erg s$^{-1}$ 

These values have been estimated in the low dispersion spectra, for an
adopted distance of 50 Mpc.

\item Velocity of the gas emitting the broad Balmer lines: \\
 $v_s$ = 7000 \kms

The freely expanding ejecta, the two thin shells associated with the
backward and leading shocks, they all contribute to the flux of the
broad Balmer lines. The maximum velocities are those of the ejecta,
thus the flux in the extended wings of the Balmer lines will come mainly from there.
Since there is probably some self-absorption, we only see the side that
is moving towards us, i.e. the blue wing. 
The main contribution to the flux of the Balmer lines is coming from
the thin shell associated with the leading shock. Its velocity can be determined therefore from the
FWZI of the core of the Balmer lines. However, as said before, due to the
self-absorption, we do not receive all of the flux emitted on the side
receding from us. Therefore, the best is to measure the velocity at
zero intensity of the bluest part of the core of the Balmer lines. The
best for that purpose is \hb\ because, while having a good
signal-to-noise ratio, it does not have the very broad
and weak tail that is present in \ha\ (see section~\ref{sec:profiles}).
We must emphasize that this value is rather qualitative. In order to
have more exact figures, one would need a more detailed modelisation -
which is beyond the scope of this paper.

\item The luminosity in the emission component of the narrow Balmer lines: \\
$L^N_{H\alpha}=4.0 \times 10^{38}$ \ergs\\
$L^N_{H\beta}= 1.1 \times 10^{38}$ \ergs

These values are measured from the echelle spectra. In fact, the true
values must be certainly larger since the blue side of the line is in
absorption. To account for this
and possible obscuration effects we have multiplied the observed
emission luminosities by a factor 2.
The \ha\ line is more intense and has better signal-to-noise ratio, but
it may be collisionally enhanced. The \hb\ line is not affected by collisional
effects but it is affected by reddening, and is weaker, which means that its
measured flux is more uncertain. 

\item Velocity of the CSM: \\
 $v_w$ = 160 \kms

It is measured from the blue wing of the absorption line in the
P~Cygni profile, in both \ha\ and \hb\ (see Fig.~\ref{fig:PCvel}). This value is larger
than the velocity of 90 \kms\ obtained for SN~1997ab (Salamanca et al. 1998).

\item Temperature of the CSM:

The coronal lines [Fe{\smc X}] 6374 \AA, [Fe{\smc XI}] 7892 \AA\ and [Fe{\smc XIV}] 5303
\AA\ indicate a medium with a high temperature, \si \E{6} K. On the
other hand, the lines [O{\smc III}] 4959,5007 \AA\AA, [N{\smc II}] 5755 \AA, \hb\ or \ha\
are formed in a somewhat cooler material, \lapprox few \ET{5} K.

\item Density of the CSM:
 
We can
impose constraints on its the density based on the ratio of the
narrow lines, [O{\smc III}] 5007 and [O{\smc III}] 4363. For high densities the
following relationship applies (Osterbrock 1989): 
\[
\frac{I_{5959} + I_{5007}}{I_{4363}} \sim \frac{7.73\times e^{\frac{32.29\times10^4}{T}}}
{1 + 4.5\times10^{-4}\frac{N_e}{\sqrt{T}}}
\]
The ratio of intensities in our case is 0.96. If we assume T to be on
the range $10^5$ -  $10^7$ K,
the density of the CSM is between 1\ET{8} and 5\ET{7}\cc.

Note as well, that the presence of [N{\smc II}] 5755 \AA\ and the
absence of [N{\smc II}] 6583 \AA\ and of [S{\smc II}] 6716,6732 \AA\
indicate that the latter are 
collisionally suppressed and therefore the CSM must have a high density.
\end{description}

\subsection{Origin of the CSM} 

The origin of such a dense material around SN~1997eg is most probably
related to a wind from the progenitor star. However, there 
are two possibilities to explain the velocity of expansion of the CSM
and its high density. 

The first, (proposed for SN~1997ab, Salamanca et al. 1998) is that this material
is entirely due to a wind of the progenitor star shortly before it exploded as a supernova.
The velocity and density of the CSM are those of the wind. This
constrains the possible candidates that are able to explode as Type
IIn SN. 
From the analysis of the spectra we can
deduce the physical conditions of this CSM, as well as the mass loss
rate of the progenitor star. 
Following the same analysis as for SN~1997ab (Salamanca et al. 1998),
we deduce that the mass loss rate of the progenitor star was \.M = 8.3
\ET{-3} \msunyr, and that the radius of the shock is of the order of
few times \E{15} cm.
We must draw the attention to the fact that in this case the density
profile of the CSM declines as r$^{-2}$$\!$.

Another  possibility is that the expansion of the CSM is
the direct consequence of the supernova explosion, and more precisely, of
the pre-ionisation of the CSM by the leading shock.
In this case the CSM is created through a ``normal''
wind in a high pressure interstellar medium (ISM). 
If the progenitor star was in a high pressure ISM, then the wind
will expand until it reaches the so-called ``stagnation point'', where it
will stop expanding and will accumulate creating a very dense shell
of constant density. When the star explodes as a supernova, the ejecta first 
expands through the progenitor
wind until it reaches the dense shell. At that moment, the phenomena of catastrophic 
cooling occurs and a luminous remnant is produced (see section 4).
The flux produced at that moment will pre-ionize\footnote{This is, prior to the ejecta
-dense CSM collision.} the CSM that
will acquire a temperature of \E{6-7} K and a thermal
velocity $v_{th} \approx c_s$ where $c_s$ is the velocity of the sound.   
Therefore, the pressure of the CSM gas will not longer be equal to
the pressure of the ISM and it will start to expand with $v_{exp} \approx
v_{th} \approx c_s$, until it reaches a new stagnation point.
For the range of temperatures of the CSM (\si few$\times$\E{5} to
few$\times$\E{6} K) the velocity of the sound is \si 55 to 170 \kms$\!$.

To distinguish both cases, we called the free expanding pre-SN wind
``Circumstellar Wind'' or CSW, and the second ``Circumstellar Shell''
or CSSh.
Note however, that these are two {\it extreme} cases, and that a
combination of both 
is not only possible, but probable.
A very simple geometry of the different layers is illustrated on Fig.~\ref{fig:circulos}. 

\subsection{Comparison with previous models}

We can compare the measured luminosities of the \ha\ and \hb\
emission lines (broad and narrow) with the theoretical values obtained
with the model of Terlevich et al. (1992, 1995). These values vary as a
function of time, which is measured in units of the time at which the forward
shock enters the radiative phase, t$_{sg}$. Assuming free-free
cooling, t$_{sg}$ is given by the expression (Shull 1980, Wheeler
et al. 1980):
\begin{equation}
t_{sg} = 230 \zp{ \frac{E}{10^{51} \,\,\rm{erg}}}^{1/8} 
\zp{  \frac{n_{csm}}{10^7 \,\, \rm{cm}^{-3}}}^{-3/4} \,\,\,\,\,\,\,\, \rm{days} 
\end{equation}

\noindent where E is the initial kinetic energy of the supernova
explosion and $n_{csm}$ is the density of the circumstellar material.

The luminosities of the narrow Balmer lines also
depend on the mass of the CSM. In Tab.~\ref{tab:rjt} we give the values for 10 \msun\
and 20 \msun. The density of the CSM is assumed constant and equal to \E{7} \cc.

\begin{table*}
  \begin{minipage}{100mm}
\caption{Luminosities of the Balmer broad  (\ha\ luminosity and
  Balmer decrement) and narrow (idem) emission lines, calculated with
  the canonical model of Terlevich et al. (1992, 1995). The density of
  the CSM is constant and equal to \E{7} \cc, and the initial energy
  of the supernova explosion is equal to 1\ET{51} erg.}
\label{tab:rjt}
\begin{tabular}{lcccccc}
\hline
t/t$_{sg}$ &  \ha Broad &  Balmer dec & \multicolumn{2}{c}{\ha Narrow} &
\multicolumn{2}{c}{Balmer dec. N}  \\
         &            &   20 \msun  & 10 \msun &  20 \msun& 10\msun \\
\hline
\hline

 1   &  41.38  &   3.7  &   39.14  & 38.76  &   5.5 &  4.9  \\
                                                     
 4   &  41.22  &   6.9  &   40.12  & 39.96  &   6.3 &  4.8  \\
                                                      
 8   &  40.87  &  14.4  &   39.78  &   -     &   5.0 &   -    \\
                                                     
16   &  40.24  &  21.7  &   39.37  &   -     &   4.0 &   -     \\
\hline
\end{tabular}
\end{minipage}
\end{table*}

If we assume that the density of the CSM is \si 1\ET{8} - 5\ET{7} \cc, then t$_{sg}$ $\approx$ 41 - 69 
days (for
E=1\ET{51} erg). On the other hand, since the SN was not seen 4 months 
before the discovery, t$_{sg}$ cannot be greater than \si 120 days.
These short values of t$_{sg}$  imply that if the CSM is in a detached
layer, it cannot be too far away from the SN.
It means as well that the value of t/t$_{sg}$ when SN~1997eg was observed was between 3 and 5.
If we now compare the values given in Table~\ref{tab:rjt} for
t/t$_{sg}$ with the measured values of the luminosities given in
section~\ref{sec:input}, we conclude that the mass of the CSM must be
around 10 \msun. 
However, these values are only indicative, and a more detailed modelization of
SN~1997eg remains to be done.

\subsection{Profile of the Narrow P Cygni}

In both cases (CSW or CSSh) the dense material in front 
of the shock will produce a P Cygni profile. The narrrow absorption line 
is produced in a a narrow ring 
where the projected velocities of the shock and CSM are
equal. Moreover, it will be
displaced to the blue, because we only see  the gas that is
coming towards us. And it will be a narrow profile because the range
of velocities observed is small (see Fig.~\ref{fig:PCvel}).
However, depending on the case, CSW or CSSh, the details of it
will be different. The evolution of it will be as well different. In
the CSW case, the velocity of the wind should be already \si 160 \kms\ when
the supernova exploded. Therefore, the P Cygni profile should stay pretty much constant,
In the CSSh case, the initial velocity of the CSM is zero, and it is
accelerated to 160 \kms\ as the rarefaction wave crosses it. This
should be reflected in an evolution with time of the P~Cygni profile.

The emission comes from two zones (E1 and E2 in
Fig.~\ref{fig:circulos}). From the gas coming towards us, we will see a
blue-shifted emission. Since the projected velocity will be larger,
this feature will be more blue-shifted than the absorption line.
If the gas moving away from us is not totally obscured, we may see
another redshifted emission. 
From the zone denominated E2 we see gas coming to and from us, and
thus the resultant profile will
be centered at the rest frame. However, the blue side will be masked
by the absorption. 

Of course, depending on the relative width of the different layers or
shells and of their optical depth, the resulting P Cygni profile will
be different. A detailed study for the CSW case is to be found in Cid-Fernandes (1999).
For the CSSh case, this needs to be done.

\section{Discussion: Circumstellar Wind or Circumstellar Shell?}

After a certain time, when the rarefaction
wave has crossed all the Circumstellar Material, both cases - CSW and
CSSh - will be undistinguible. A good observational test would be to
monitor with high resolution spectroscopy the same Type IIn supernova
at different epochs, starting as close as possible to the explosion
date. If the the P Cygni profile remains constant, then this will
point to a CSW, otherwise to a CSSh.
In other words, we need high {\it spectral and temporal} resolution
spectra to be able to deduce which is the density profile of the CSM.
Furthermore, the other narrow lines associated with the CSM should
give us a good deal of information. Does the ratio of Oxygen lines [O
{\smc III}] 4363/5007 remains constant with time? Are other coronal lines
appearing at a later epoch? Is there any evolution in their luminosities
or ratios? If yes, is this evolution the
same as in the continuum or as in the broad Balmer lines? 

Besides, we must not forget that Type IIn SN are not only very energetic in
the visible, but as well in X-rays (Fabian \& Terlevich 1996, Lewin,
Zimmermann, Aschenbach 1995) or Radio (Van Dyk et al 1993). In order to  
have a global and complete picture, we should
have multi-monitoring campaigns in the same way as is done with AGN
(e.g. Peterson \& Wandel 2000, Edelson et al. 2000, Nandra et al 1998).
Only then we would have enough observational constrains
for the theoretical models. In particular, we should be able to deduce
observationally the physical parameters of the CSM. This is so
important because such CSM gives us important clues about the
progenitor star and its environment.
In the CSW case, this wind must be extremely dense. Its
existence gives us direct clues about the nature of the progenitor
star. In the CSSh case, the wind could be a ``normal'' RSG wind, but
the progenitor must be embedded into a very high pressure medium. In
such case, the progenitor wind gives us information about the regions
where the star has formed and lived.                 

If we assume the CSW case, from the values of \.M and $v_w$ inferred we can constrain
possible candidates for progenitors. Let's examine three of them,
known to have important mass losses: O stars, Wolf-Rayet stars and Luminous Blue
Variables (also called S Dor variables). 
 
O-stars have winds with \.M \si \E{-5} \msunyr\ and
velocities of typically 10 \kms. Recent calculation by Panagia \& Bono
(2000) show that massive stars can have higher mass loss rates, up to
\si \E{-3} \msunyr\ and with velocities up to 50 \kms. However, this is
out of the range of values measured for SN~1997ab or 1997eg.

Wolf-Rayet stars have much faster and massive winds: the velocity of
their winds is of the order of few\ET{3} \kms\ and the mass loss
rates are \si 2\ET{-5} \msunyr\ (Abbott \& Conti 1987). Again, those
numbers are different for what we find in SN~1997eg. However, they are
more close to the values derived for less luminous Type IIn supernovae (see
section~\ref{sec:othercases}).

Luminous Blue Variables stars (LBV) (see Humphreys \& Davidson
1994 for a review) can experience very high mass losses in a very
short time, during giant eruptions. 
The maximum mass lost rates deduced for LBV are in
the order of \.M \si $10^{-3} - 10^{-4}$ \msun yr$^{-1}$. 
For some cases, (e.g. $\eta$ Carinae) the mass loss rate could reach values
up to \si 10$^{-1}$ \msun yr$^{-1}$ (Davidson 1989). 
This ``super-wind'' has a velocity within
the values measured for the CSM \si 40 \kms\ up to 700
\kms. The density of the nebula created by this wind is also in
the range of densities measured for the material around Typw IIn SN : n$_e$ \si
10$^3$ to 10$^7$ \cc\ (Stahl 1989). 

Obviously, the issue of the nature of the progenitors of Type IIn
deserves more thorough investigation, but the possibility that 
they are post LBV stars opens an interesting question.

%{\it Massive winds, yes ... but which environment?}

In we place ourselves in the CSSh case, then this means that the
progenitor star was inside a very high pressure ambient medium. Where
can we find such a medium?  The most obvious answer is in star forming
regions, where the wind of many massive stars and supernova
explosions can create such high ambient pressure medium. One may ask
then, why we do not see more of this Type IIn supernovae on Starburst
galaxies? The hypothesis, proposed by Terlevich et al. (1992, 1995), is that we {\it do see}
them, and we call them Active Galactic Nuclei (AGN).

\subsection{Other cases of narrow P Cygni profiles}
\label{sec:othercases}

It is becoming clear that the narrow P Cygni of SN~1997eg is
not an exception, but rather a common feature of Type IIn
supernovae, at least in the initial stages of their evolution. The
cases known until now are SN~1995G (Filippenko \& Schlegel 1995,
Turatto private communication), SN~1997ab 
(Salamanca et al. 1998) and SN1998S (Fassia et al 2001). The
expansion velocity 
of the CSM deduced for SN~1997ab was $v_w \sim 90$ \kms and for
SN~1998S, $v_w \sim 45$ \kms.
These values seem to indicated that there is a range in the velocities
of the dense CSM around Type IIn Supernovae.

Moreover, there are less luminous Type IIn supernovae (classified as
Type IId by Benetti et al. 1998),
which also have P Cygni
profiles atop the broad emission lines. The difference with the more
luminous SN IIn are that the P Cygni lines are somewhat broader (of about 1000
\kms) and that the 
broad emission lines are narrower, and less luminous. This can be
explained by a lower CSM density, n \si 10$^5$ \cc, and by a lower mass
loss rate of the progenitor star, \.M \si 10$^{-4}$ (assuming the CSW hypothesis). A good
example is SN~1994aj (Benetti et al. 1998). 

Another example of detection of the dense CSM is SN~1978K (Chu et al. 1999).

\section{Conclusions}

There is strong observational evidence 
for the existence of a dense and hot CSM in Type
IIn SN, particularly in SN~1997eg. Such dense environment was
``necessary'' from the theoretical point of view, because the physical
explanation invokes the presence of radiative shocks, produced
via the interaction of the SN ejecta with dense material.
Echelle spectrum of SN~1997eg shows a very narrow P Cygni line atop the
broad emission on \ha\ and \hb. This feature seems to be common in Type
IIn SN in their early stages, and points to either a massive and slow wind
of the progenitor just prior to the explosion or to a wind in a high
pressure medium, as its origin.

However, even though the data presented here answer many questions
concerning such CSM, it opens even more, which can only be
addressed with more data. Instead of concentrating in small details
of SN IIn, long-term, multi-wavelength monitoring of such objects
would be the right thing to do (e.g. Aretxaga et al. 1998) in a similar
way as is done for AGN (see for example Peterson \& Wandel 2000).

\section*{Acknowledgements} 

I.S is deeply obliged to Maureen van der Berg for her help in calibrating the
echelle spectra and to Dr. Arnout van Genderen for his useful comments
on an earlier draft of this paper. We are as well thankful to the anonymous 
referee for his/her comments which have improved the clarity of the paper.

This research has been supported by a Marie Curie Fellowship of the
European Community programme ``Training and Mobility of Researches''
awarded to I. Salamanca (proposal Nr. ERB4001GT974289).

This project has been supported by the European Commission through the
Activity ``Access to Large-Scale Facilities'' within the Program
``Training and Mobility of Researches'', awarded to the Instituto de
Astrof\'{\i}sica de Canarias to fund European Astronomers' access to
its Roque de los Muchachos and Teide Observatories (European Northern
Observatory), in the Canary Islands.

\bsp 
 
\label{lastpage} 


\begin{thebibliography}{} % Note the pair of empty curly braces!
\bibitem{} Abbott D.C., Conti P.S., 1987, ARA\&A, 25, 113
\bibitem{} Aretxaga I., Benetti S., Terlevich R.J., Fabian A.C., Cappellaro E., Turatto M.,
 della Valle M., 1998, MNRAS, 309, 343
\bibitem{} Benetti S., Cappellaro E., Danzinger I.J., Turatto M., Patat F., 
Della Valle M., 1998, MNRAS, 294, 448
\bibitem{} Chevalier R.A., Fransson C. 1994, ApJ, 420, 268
\bibitem{} Chu Y.-H., Caulet A., Montes M.J., Panagia N., van Dyk S. D., Weiler K. W., 1999, ApJ, 512, L51
\bibitem{} Chugai N.N., 1991, MNRAS, 250, 513
\bibitem{} Cid-Fernandes R., 1999, MNRAS, 305, 602
\bibitem{} Davidson K., 1989, in Davidson K., Moffat A.F.J.,  Lamers H.J.G.L.M., eds., 
Physics of Luminous Blue Variables, Kluwer Academic Publ., p. 101
\bibitem{} Edelson R., Koratkar A., Nandra K. et al., 2000, ApJ, 534, 180
\bibitem{} Fabian A.C., Terlevich R.J., 1996, MNRAS, 280, L5
\bibitem{} Fassia A., Meikle W.P.S., Chugai N. et al., 2001, MNRAS, 325, 907
\bibitem{} Filippenko A.V., Barth A.J., 1997, IAU Circular no. 6794
\bibitem{} Filippenko A.V., Schlegel D., 1995, IAU Circular no. 6139
\bibitem{} Ho L.C., Filippenko A.V., Sargent W.L.W., 1997, ApJS, 112, 315
\bibitem{} Humphreys R.M., Davidson K., 1994, PASP, 106, 1025
\bibitem{} Lacey C.K., Weiler K.W., 1998, IAU Circular no. 7068
\bibitem{} Lewin W.H.G., Zimmermann H.-U., Aschenbach B., 1995, IAU Circ no. 6445
\bibitem{} Meikle P., Geballe T., 1998, UKIRT Newsletter vol 3
\bibitem{} Nandra K., Clavel J., Edelson R.A., George I.M., Malkan M.A., 
Mushotzky R.F., Peterson B.M., Turner T.J., 1998 ApJ, 505, 594
\bibitem{} Nakano S., Masakatsu A., 1997, IAU Circular no. 6790
\bibitem{} Osterbrock D.E., 1989, Astrophysics of Gaseous Nebulae
  and Active Galactic Nuclei. Mill Valley, CA, University Science Books
\bibitem{} Panagia N., Bono G., 2000, in Livio M., Panagia N., Sahu K., eds., 
STScI May Symposium, ``The Largest Explosions since the Big-Bang: Supernovae and Gamma Ray Bursts'', 
CUP-Cambridge, in press.
\bibitem{} Peterson B.M., Wandel A., 2000 ApJ, 540, L13
\bibitem{} Salamanca I., Cid-Fernandes R., Tenorio-Tagle G., Telles E., Terlevich R.J., 
Mu\~noz-Tu\~non C., 1998, MNRAS, 300, L17
\bibitem{} Shull M., 1980, ApJ, 237, 769 
\bibitem{} Stahl O., 1989, in Davidson K., Moffat A.F.J., Lamers H.J.G.L.M. eds., 
Physics of Luminous Blue Variables. Kluwer Academic Publ., p. 149
\bibitem{} Terlevich R.J. Tenorio-Tagle G., Franco J., Melnick J., 1992, MNRAS, 255, 713
\bibitem{} Terlevich R.J., Tenorio-Tagle G., Rozyczka M., Franco J., Melnick J., 1995, MNRAS, 272, 198
\bibitem{} Van Dyk S.D., Weiler K.W., Sramek R.A., Panagia, N., 1993 ApJ, 419, L69
\bibitem{} Wheeler J.C., Mazurek T.J., Sivaramakrishnan A., 1980, ApJ, 237, 781
\end{thebibliography}
\end{document}